\documentstyle[emulateapj]{article}

\def\lsim{\mathrel{\rlap{\lower 4pt \hbox{\hskip 1pt $\sim$}}\raise 1pt \hbox
        {$<$}}}
\def\gsim{\mathrel{\rlap{\lower 4pt \hbox{\hskip 1pt $\sim$}}\raise 1pt \hbox
        {$>$}}}

\lefthead{Nakamura et al.}
\righthead{Explosive Nucleosynthesis
in Hyper-Energetic Supernovae, Hypernovae}

\begin{document}

\submitted{Accepted for publication
in the Astrophysical Journal (13 March 2001)}

\title{Explosive Nucleosynthesis in Hypernovae}

\author{Takayoshi Nakamura\altaffilmark{1}, Hideyuki Umeda\altaffilmark{2, 8},
Koichi Iwamoto\altaffilmark{3}, Ken'ichi Nomoto\altaffilmark{4, 8},
Masa-aki Hashimoto\altaffilmark{5},
W. Raphael Hix\altaffilmark{6}, 
and Friedrich-Karl Thielemann\altaffilmark{7}}

\altaffiltext{1}{Department of Astronomy, School of Science,
University of Tokyo, Tokyo, Japan; nakamura@astron.s.u-tokyo.ac.jp}
\altaffiltext{2}{Department of Astronomy, School of Science,
University of Tokyo, Tokyo, Japan; umeda@astron.s.u-tokyo.ac.jp}
\altaffiltext{3}{Department of Physics, College of Science and Technology,
Nihon University, Tokyo, Japan; iwamoto@etoile.phys.cst.nihon-u.ac.jp}
\altaffiltext{4}{Department of Astronomy, School of Science,
University of Tokyo, Tokyo, Japan; nomoto@astron.s.u-tokyo.ac.jp}
\altaffiltext{5}{Department of Physics, School of Sciences,
University of Kyushu, Fukuoka, Japan; hashi@gemini.rc.kyushu-u.ac.jp}
\altaffiltext{6}{Department of Physics and Astronomy, University of 
Tennessee, Knoxville, TN 37996-1200 
and Physics Division, Oak Ridge National Laboratory, Oak Ridge, TN 
37831-6354
and Joint Institute for Heavy Ion Research, Oak Ridge National Laboratory,
Oak Ridge, TN 37831-6374, USA; raph@mail.phy.ornl.gov}
\altaffiltext{7}{Department f\"ur Physik und Astronomie,
Universit\"at Basel, Switzerland; fkt@quasar.physik.unibas.ch}
\altaffiltext{8}{Research Center for the Early Universe, School of Science,
University of Tokyo, Tokyo, Japan}

\begin{abstract}

We examine the characteristics of nucleosynthesis in 'hypernovae',
i.e., supernovae with very large explosion energies ($ \gsim 10^{52} $
ergs).  We carry out detailed nucleosynthesis calculations for these
energetic explosions and compare the yields with those of ordinary
core-collapse supernovae.  We find that both complete and incomplete
Si-burning takes place over more extended, lower density regions, so
that the alpha-rich freezeout is enhanced and produces more Ti in
comparison with ordinary supernova nucleosynthesis.  In addition,
oxygen and carbon burning takes place in more extended, lower density
regions than in ordinary supernovae.  Therefore, the fuel elements O,
C, Al are less abundant while a larger amount of Si, S, Ar, and Ca
("Si") are synthesized by oxygen burning; this leads to larger ratios
of "Si"/O in the ejecta.  Enhancement of the mass ratio between
complete and incomplete Si-burning regions in the ejecta may explain
the abundance ratios among iron-peak elements in metal-poor stars.
Also the enhanced "Si"/O ratio may explain the abundance ratios
observed in star burst galaxies.  We also discuss other implications
of enhanced [Ti/Fe] and [Fe/O] for Galactic chemical evolution and the
abundances of low mass black hole binaries.

\end{abstract}


\section{Introduction}\label{sec:intro}

Massive stars in the range of 8 to $\sim$ 150$M_\odot$ undergo
core-collapse at the end of their evolution and become Type II and
Ib/c supernovae unless the entire star collapses into a black hole
with no mass ejection (e.g., Arnett 1996).  These Type II and Ib/c
supernovae (as well as Type Ia supernovae) release large explosion
energies and eject explosive nucleosynthesis materials, thus having
strong dynamical, thermal, and chemical influences on the evolution of
interstellar matter and galaxies.  Therefore, the explosion energies
of core-collapse supernovae are fundamentally important quantities,
and an estimate of $E \sim 1\times 10^{51}$ ergs has often been used
in calculating nucleosynthesis (e.g., Woosley \& Weaver 1995;
Thielemann et al. 1996) and the impact on the interstellar medium.
(In the present paper, we use the explosion energy $E$ for the final
kinetic energy of explosion.)
A good example of this estimate
is SN1987A in the Large Magellanic Cloud, whose energy is
estimated to be $E = (1.0$ - 1.5) $\times$ $10^{51}$ ergs from its
early light curve (Shigeyama et al. 1987, 1988; Woosley et al. 1988;
Arnett et al. 1989; Nakamura et al. 1998; Blinnikov et al. 2000).

SN1998bw called into question the applicability of the above energy
estimate for all core-collapse supernovae.  This supernova was
discovered in the error box of the gamma-ray burst GRB980425 (Soffitta
et al. 1998; Lidman et al. 1998; Galama et al. 1998).  Although the
association of supernovae with gamma-ray bursts is still the subject
of controversy (e.g., Wang \& Wheeler 1998; Norris et al. 1999),
SN1998bw is a quite unusual supernova in its radio and optical
properties (Galama et al. 1998; Iwamoto et al. 1998; Kulkarni et
al. 1998).  It is one of the most luminous supernovae at radio
wavelengths and its peak luminosity was attained exceptionally early
in comparison with other radio supernovae (Kulkarni et al. 1998).
From the optical spectra SN1998bw is classified as a Type Ic supernova
(SN Ic), but it shows unusually broad spectral features.  The modeling
of the early light curve and spectra leads us to conclude that it was
an explosion of a $\sim$ 14 $M_\odot$ C + O star with the explosion
energy $E$ = 3 - 6 $ \times$ $10^{52}$ ergs, which is about thirty
times larger than that of a canonical supernova (Iwamoto et al. 1998;
Woosley, Eastman \& Schmidt 1999; Branch 2000; Nakamura et al.
2000, 2001).  The amount of $^{56}$Ni ejected from SN1998bw is found to be
$M(^{56}$Ni) $\simeq$ 0.4 - 0.7 $M_\odot$ (Nakamura et al. 2000, 2001;
Sollerman et al. 2000), which is about 10 times larger than the
0.07$M_\odot$ produced in SN1987A, a typical value for core-collapse
supernovae.

We have used the term 'hypernova' to describe such an
extremely energetic
supernova with $E \gsim 10^{52}$ ergs (Nomoto et al. 2000).
In terms of explosion energies, pair-instability supernovae of
150 - 200 $M_\odot$ stars also explode with $E \gsim 10^{52}$ ergs,
thus being called hypernovae
(e.g.,Ober et al. 1983; Bond et al. 1984; Stringfellow \& Woosley
1988).
In the present paper, we concentrate on nucleosynthesis in
core-collapse hypernovae.

Recently, other hypernovae candidates have been recognized.  SN1997ef and
SN1998ey are also classified as SNe Ic, and show very broad spectral
features similar to SN1998bw (Garnavich et al. 1997, 1998).  The
spectra and the light curve of SN1997ef have been well simulated by
the explosion of a 10$M_\odot$ C+O star with $E = 1.0 \pm 0.2$
$\times$ $10^{52}$ ergs and $M(^{56}$Ni) $\sim$ 0.15 $M_\odot$
(Iwamoto et al. 2000; Mazzali, Iwamoto, \& Nomoto 2000).  SN1997cy is
classified as a SN IIn and unusually bright (Germany et al. 2000;
Turatto et al. 2000). Its light curve has been simulated by a
circumstellar interaction model which requires $E \sim$ 5 $\times$
$10^{52}$ ergs (Turatto et al. 2000).  The spectral similarity of
SN1999E to SN1997cy (Cappellaro et al. 1999) would suggest that SN1999E
is also a hypernova.  Note that all of these estimates of $E$ assume a
spherically symmetric event.

In this paper, we explore nucleosynthesis in such energetic
core-collapse induced
supernovae, the systematic study of which has not yet been done.
In \S \ref{sec:98bw}, we briefly describe the argument that the
progenitor mass $M$ and explosion energy $E$ are very large from the
observation of SN1998bw.  In \S \ref{sec:model} and \S
\ref{sec:depend}, the characteristics of nucleosynthesis in hypernovae
are investigated with detailed nucleosynthesis calculations and
compared with nucleosynthesis in canonical supernovae.  We also
discuss possible effects on the Galactic chemical evolution and on the
abundances in metal-poor stars in \S \ref{sec:diss}. A summary is
given in section \ref{sec:summary}.

\section{The progenitor of SN1998bw}\label{sec:98bw}

In this section, we will use SN1998bw, the first discovered hypernova,
to describe how one can constrain the progenitor mass $M$ and the
explosion energy $E$ from the observed light curve width and the peak
luminosity by using the relation between $M$, $E$ and the synthesized
$^{56}$Ni mass $M$($^{56}$Ni).  The light curve of SN1998bw (Galama et
al. 1998) showed a very early rise and reached the peak luminosity at
$\sim 18$ days after the explosion before declining exponentially with
time.  This decline clearly indicates that the light curve is powered by the
radioactive decay of $^{56}$Ni and $^{56}$Co as in usual SNe.  The
distance to the host galaxy ESO184-G82 is estimated to be $\sim 37.8$
Mpc (Patat et al. 2001).  From this distance, the peak absolute
luminosity is estimated to be $\sim 1 \times$ $10^{43}$ ergs s$^{-1}$,
which is about ten times larger than previous SNe II or SNe Ib/c
(e.g., SN1994I; Nomoto et al. 1994; Iwamoto et al. 1994), and as
bright as an average SN Ia.  The ejected $^{56}$Ni mass necessary to
achieve such a large luminosity is estimated to be 0.4 - 0.7$M_\odot$,
which is also about ten times larger than those of typical
core-collapse SNe.

The peak width of the light curve, $\tau_{\rm peak}$, is approximately
related to the mass of the ejecta $M_{\rm ej}$ and $E$ as follows.
\begin{eqnarray} 
\tau_{\rm peak} \sim \sqrt{\tau_{\rm diff} \cdot \tau_{\rm dyn}} \sim
( \frac{\kappa}{c} )^{1/2} M_{\rm ej}^{3/4}
E^{-1/4},\label{eq:taupeak}
\end{eqnarray} 
where $\kappa$ is the opacity and $c$ is the speed of light.  This
relation is derived from multiplying the time scale of photon
diffusion $\tau_{\rm diff} \sim \kappa \rho R^2 /c$ with the dynamical
time scale $\tau_{\rm dyn} \sim R/v$ (Arnett 1982, 1996).  This means
that we can determine only $M_{\rm ej}^{3}/E$ by the light curve
fitting.  However, the information about 
the photospheric velocity can be
used to break the degeneracy.  In fact, Iwamoto et al. (1998) and
Nakamura et al. (2000, 2001) used the photospheric velocity to constrain
$M_{\rm ej}^{1/2}/E^{1/2}$ and determined $M_{\rm ej}$ and $E$ as
$M_{\rm ej} \sim 10 $ - $ 11M_{\odot}$ and $E \sim 3$ - $6$ $\times$
$10^{52}$ ergs. From this ejecta mass and spectral type (Ic), they
also concluded that the progenitor of SN1998bw was a C + O star with
mass of $\sim$ 14$M_{\odot}$, corresponding to an initial
main-sequence mass $\sim$ 40$M_{\odot}$.  This is relatively large
compared with determinations of 
$\sim$ 20$M_{\odot}$ for SN1987A (Arnett et al. 1989;
Shigeyama \& Nomoto 1990), $\sim$ 13$M_{\odot}$ for SN1993J (Nomoto et
al. 1993a; Woosley et al. 1994), and $\sim$ 14$M_{\odot}$ for SN1994I
(Nomoto et al. 1994; Iwamoto et al. 1994; Young et al. 1995).

We can also exclude conventional masses of progenitors without
using the information from photospheric velocities, but instead utilizing the
peak luminosity, i.e., $M$($^{56}$Ni).
The products of explosive burning are largely determined   by the
maximum temperature behind the shock, $T_{\rm s}$.
Material which experiences $T_{\rm s} > 5$ $\times$ $10^{9}$ K
undergoes complete Si-burning,
forming predominantly $^{56}$Ni.  We can estimate the
radius of the sphere in which $^{56}$Ni is dominantly produced as
\begin{eqnarray} 
R_{\rm Ni} \sim 3700 \hspace{2mm} (E/10^{51} {\rm ergs})^{1/3}
\hspace{3mm} {\rm km}, \label{eq:rni}
\end{eqnarray}
which is derived from $E = 4\pi/3R_{\rm Ni}^{3} a T_{\rm s}^{4}$ with
$T_{\rm s} = 5$ $\times$ $10^{9}$K (e.g., Thielemann et al. 1996).
Note that equation (\ref{eq:taupeak}) indicates that smaller $M_{\rm
ej}$ correspond to smaller $E$ for a given $\tau_{\rm peak}$, while
equation (\ref{eq:rni}) relates a smaller $E$ to smaller $R_{\rm Ni}$.
Thus, combining these two equations, we find that $R_{\rm Ni}$ is
smaller for smaller $M_{\rm ej}$.  In other words, a small mass
progenitor may not produce a significantly large amount of $^{56}$Ni.

\begin{figure}
\epsscale{.7}
\plotone{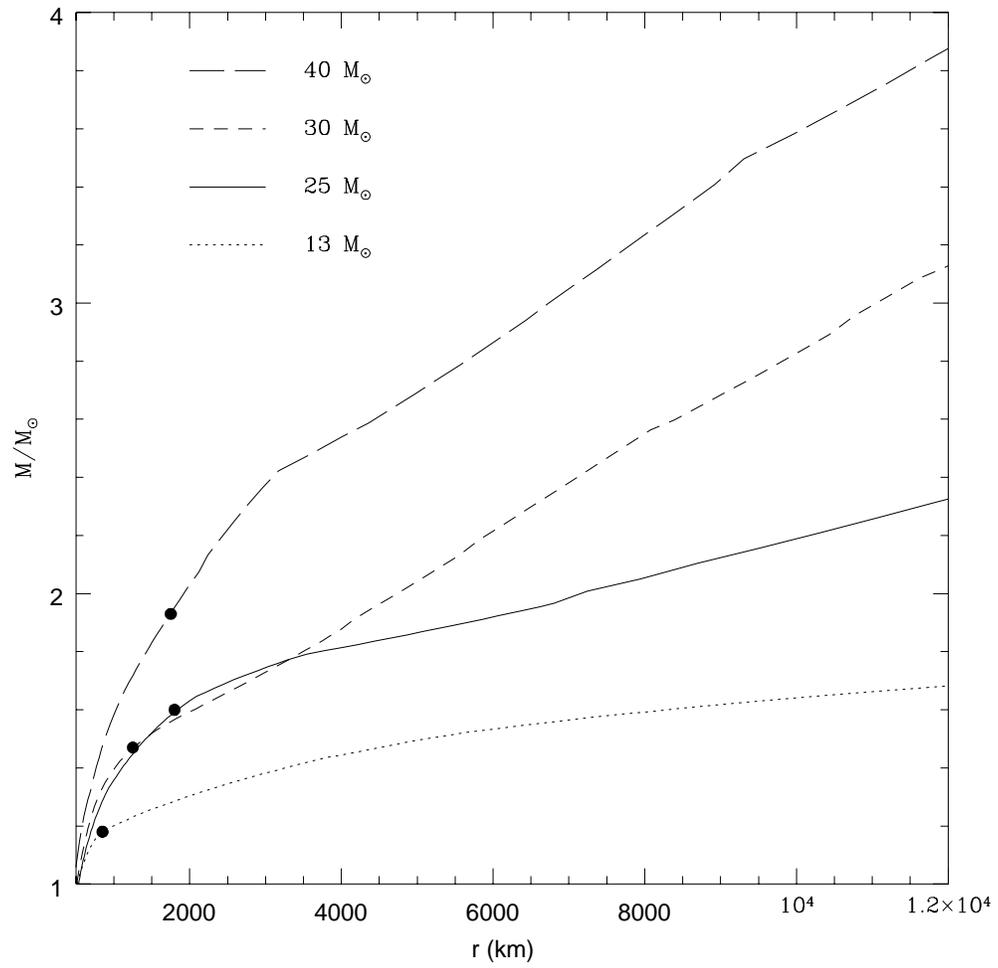}
\caption{The mass enclosed within a radius $r$ of the pre-collapse stars.
Filled circles locate the position of the pre-collapse Fe core.
\label{fig:mvsr}}
\end{figure}

Figure \ref{fig:mvsr} shows the mass enclosed within a radius $r$ of
the pre-collapse stars for stellar masses of $M$ = 13 - 40 $M_\odot$
(Nomoto et al. 1993b).  We can see that the stellar structure actually
provides the above relation, i.e., smaller $M_{\rm ej}$ corresponding
to a smaller amount of $R_{\rm Ni}$.  A less massive star contains a
smaller amount of mass in the same radius.  Therefore, a less massive
star ejects a smaller amount of $^{56}$Ni than a more massive star,
when they are subjected to explosions of the same energy.

For SN1998bw, 0.4 - 0.7 $M_\odot$ $^{56}$Ni is ejected.  Since the
mass of a collapsed star (neutron star or black hole) is at least as
large as the mass of the pre-collapse Fe core, being 1.2 - 1.9
$M_\odot$ as indicated in Figure \ref{fig:mvsr}, the mass contained
within $R_{\rm Ni}$ should exceed $\sim$ 1.6 - 2.3 $M_\odot$.  For
example, consider an explosion of a 6$M_\odot$ C+O star, corresponding
to a main-sequence mass $\sim 25 M_\odot$.  If such a C+O star
explodes, the ejecta mass becomes $M_{\rm ej} \lsim$ 4.4 $M_\odot$,
which, along with the light curve equation (\ref{eq:taupeak}), yields
$E \lsim 2$ $\times$ $10^{51}$ ergs.  Substituting $E \lsim 2$
$\times$ $10^{51}$ ergs for equation (\ref{eq:rni}), we obtain $R_{\rm
Ni} \lsim 4700$km. Within this $R_{\rm Ni}$, only 1.8$M_\odot$ is
enclosed (Figure \ref{fig:mvsr}), so that only 0.2$M_\odot$ $^{56}$Ni
can be ejected.  From this, we exclude the models less massive than
25$M_\odot$ in their main sequence because they cannot eject enough
$^{56}$Ni.  This argument above is further evidence that the
progenitor of SN1998bw is a massive star.

Since the progenitor of another hypernova candidate SN1997ef also
seems to be massive (Iwamoto et al. 2000), we consider only relatively
massive progenitor models (8 - 16 $M_\odot$ He core models) in the
following for nucleosynthesis calculations of hypernovae.

\section{Models}
\label{sec:model}

Hypernovae are characterized by explosion energies larger than $E \sim
10^{52}$ ergs.  Such an energetic stellar explosion may be associated
with the formation of a black hole as has been discussed in the
context of the GRB-SNe connection (Woosley 1993; Paczynski 1998;
Iwamoto et al. 1998; MacFadyen \& Woosley 1999).  In these models, the
gravitational energy, or the rotational energy would be released via
pair-neutrino annihilation or the Blandford-Znajek mechanism
(Blandford \& Znajek 1977).  Alternatively, large magnetic energies
are released from a possible magnetar (Nakamura 1998; Wheeler et
al. 2000).  The explosion may also be aspherical (H\"oflich, Wheeler,
\& Wang 1999; MacFadyen \& Woosley 1999; Khokhlov et al. 1999).
However, the actual explosion mechanism and the degree of asphericity
in the ejecta are still quite uncertain.  For the present paper,
therefore, we investigate nucleosynthesis in spherical explosions as
an extreme case.  In a next step, we will explore aspherical explosion
models (Maeda et al. 2000; also Nagataki et al. 1997).

Our calculations are performed in the same way as studies of supernova
nucleosynthesis (e.g., Hashimoto et al. 1989; Thielemann, Hashimoto,
\& Nomoto 1990; Hashimoto 1995; Thielemann, Nomoto, \& Hashimoto 1996;
Nakamura et al. 1999).  First, the hydrodynamical simulations are
performed with a one dimensional PPM (piecewise parabolic method) code
(Colella \& Woodward 1984), which includes a small nuclear reaction
network that contains only 13 alpha nuclei ($^{4}$He, $^{12}$C,
$^{16}$O, $^{20}$Ne, $^{24}$Mg, $^{28}$Si, $^{32}$S, $^{36}$Ar,
$^{40}$Ca, $^{44}$Ti, $^{48}$Cr, $^{52}$Fe, and $^{56}$Ni) in order to
take into account the energy release due to nuclear reactions.  We
generate a shock wave by depositing thermal energy below the mass cut,
which divides the central compact object and the ejecta.  Next,
post-processing calculations are performed at each mesh point of the
hydrodynamical model with an extended reaction network of 293 isotopes
(Hix \& Thielemann 1996, 1999) to provide precise total yields (even
for minor abundances).  The progenitor models are taken from Nomoto \&
Hashimoto (1988), Hashimoto (1995) and Nomoto et al. (1997).  We make
use of the 6$M_{\odot}$, 8$M_{\odot}$, 10$M_{\odot}$, and
16$M_{\odot}$ He core models, which correspond approximately to the
main sequence masses of 20$M_{\odot}$, 25$M_{\odot}$, 30$M_{\odot}$,
and 40$M_{\odot}$, respectively.  In order to compare nucleosynthesis
in hypernovae with ordinary supernovae and also to investigate the
dependence on the explosion energy, we study explosion energies of $E
=$ 100, 30, 10, and 1 $ \times$ $ 10^{51}$ ergs.  The mass cuts are
summarized in Table \ref{tab:masscuthn}.  We chose these mass cuts to
be as small as they can, but prevent O/Fe of the ejecta from being
much less than the solar value (\S \ref{sec:alpha}).

\section{Dependence on Explosion Energy}\label{sec:depend}

Figure \ref{fig:donehn} shows nucleosynthesis in hypernovae and
typical supernovae for $E =$ 100 (top left), 30 (top right), 10
(bottom left), and 1 (bottom right) $ \times$ $ 10^{51}$ ergs.  The
progenitor is 16$M_{\odot}$ He core model. From this figure, we note
the following characteristics of nucleosynthesis with very large
explosion energies.

\begin{enumerate}

\item The region of complete Si-burning, where $^{56}$Ni is dominantly
produced, is extended to the outer, lower density region.  How much
mass is ejected from this region depends on the mass cut.  The large
amount of $^{56}$Ni observed in hypernovae (e.g., $ \sim$ 0.4 - 0.7
$M_{\odot}$ for SN1998bw and $ \sim 0.15 M_{\odot}$ for SN1997ef)
implies that the mass cut is rather deep, so that the elements
synthesized in this region such as $^{59}$Cu, $^{63}$Zn, and $^{64}$Ge
(which decay into $^{59}$Co, $^{63}$Cu, and $^{64}$Zn, respectively)
are ejected more abundantly than in normal supernovae.  (The mass
fraction of $^{63}$Zn is so small, $\lsim 1 \times$ $ 10^{-5}$, that
$^{63}$Zn is out of range in Figure \ref{fig:donehn}.)  In the
complete Si-burning region of hypernovae, elements produced by
$\alpha$-rich freezeout are enhanced because nucleosynthesis proceeds
at lower densities than in normal supernovae (Figure \ref{fig:maxrt}).
Figure \ref{fig:donehn} clearly shows the trend that a larger amount
of $^{4}$He is left for more energetic explosions.  Hence, elements
synthesized through capturing of $\alpha$-particles, such as
$^{44}$Ti, $^{48}$Cr, and $^{64}$Ge (decaying into $^{44}$Ca,
$^{48}$Ti, and $^{64}$Zn, respectively) are more abundant.

\item More energetic explosions produce a broader incomplete
Si-burning region.  The elements produced mainly in this region such
as $^{52}$Fe, $^{55}$Co, and $^{51}$Mn (decaying into $^{52}$Cr,
$^{55}$Mn, and $^{51}$V, respectively) are synthesized more abundantly
with the larger explosion energy.

\item Oxygen and carbon burning takes place in more extended, lower
density regions for the larger explosion energy.  Therefore, the
elements O, C, Al are burned more efficiently and the abundances of
the elements in the ejecta are smaller, while a larger amount of
burning products such as Si, S, and Ar is synthesized by oxygen
burning.
\end{enumerate}

Tables \ref{doneisohn16a} - 9 summarize the
nucleosynthesis products of hypernovae (and normal supernovae)
before and after radioactive decays for various explosion energies.
Major radioactive elements are summarized in table \ref{doneisohn}.
The progenitors are He stars of
16$M_{\odot}$, 10$M_{\odot}$, 8$M_{\odot}$, and 6$M_{\odot}$.
Products from the H-rich envelope are not included.
Note that the amount of the elements produced in the
complete Si-burning region depends on the mass cut. Here we choose the
mass cuts shown in Table \ref{tab:masscuthn}.

Figure \ref{fig:vssolarhn} displays the abundances of stable isotopes
relative to their solar values for $E =$ 100 (top left), 30 (top
right), 10 (bottom left), and 1 (bottom right) $ \times$ $ 10^{51}$
ergs.  The progenitor is a 16$M_{\odot}$ He star.  The isotopic ratios
relative to $^{16}$O with respect to the solar values are shown.  The
nucleosynthesis is characterized by large abundance ratios of
intermediate mass nuclei and heavy nuclei with respect to $^{56}$Fe
for more energetic explosions, except for the elements O, C, Al which
are consumed in oxygen and carbon burning.  In particularly, the
amounts of $^{44}$Ca and $^{48}$Ti are increased significantly with
increasing explosion energy because of the lower density regions which
experience complete Si-burning through $\alpha$-rich freezeout.  To
see this more clearly, we add the isotopes and show in Figure
\ref{fig:ratiohn} their ratios to oxygen relative to the solar values
for various explosion energies.  We note that [C/O] and [Mg/O] are not
sensitive to the explosion energy because C, O, and Mg are consumed by
oxygen burning.  On the other hand, [Si/O], [S/O], [Ti/O], and [Ca/O]
are larger for larger explosion energies because Si and S are produced
in the oxygen burning region, and Ti and Ca are increased in the
enhanced $\alpha$-rich freezeout.

\section{Contribution of Hypernovae to the Galactic Chemical Evolution}
\label{sec:diss}

Because of the small number of hypernovae so far observed, their
occurrence frequency is difficult to estimate.  However, their effects
on the Galactic chemical evolution could be important because of their
production of large amounts of heavy elements.

If hypernovae occurred in the early stage of the Galactic evolution,
the abundance pattern of a hypernova may be observable in some
low-mass halo stars.  This results from the very metal-poor
environment, where the heavy elements synthesized in a single
hypernova (or a single supernova) dominate the heavy element abundance
pattern (Audouze \& Silk 1995).  It is plausible that hypernova
explosions induce star formations.  The low-mass stars produced by
this event should have the hypernova-like abundance pattern and still
exist in the Galactic halo.  The metallicity of such stars is likely
to be determined by the ratio of ejected iron mass from the relevant
hypernova to the mass of hydrogen gathered by the hypernova, which
might be in the range range of $ -4 \lsim$ [Fe/H] $\lsim -2.5$ ([A/B]
= log$_{10}$(A/B)$-$log$_{10}$(A/B)$_\odot$; Ryan et al. 1996;
Shigeyama \& Tsujimoto 1998; Nakamura et al. 1999; Argast et
al. 2000).

We also discuss the abundances of the black hole binary X-ray Nova Sco
as a possible direct indication of hypernova nucleosynthesis.

\subsection{Alpha-Elements and Ti relative to Fe}\label{sec:alpha}

First of all, the most significant feature of hypernova
nucleosynthesis is a large amount of Fe.  One hypernova can produce
2 - 10 times more Fe than normal core-collapse supernovae,
which is almost the same amount of Fe as produced in a SN Ia.  This
large iron production leads to small ratios of $\alpha$ elements over
iron in hypernovae (Figure \ref{fig:vssolarhn}).  In this connection,
the abundance pattern of the very metal-poor binary CS22873-139
([Fe/H] = $-3.4$) is interesting.  This binary has only an upper limit
to [Sr/Fe] $< -1.5$, and therefore was suggested to be a second
generation star (Nordstr\"om et al. 2000; Spite et al. 2000).  The
interesting pattern is that this binary shows almost solar Mg/Fe and
Ca/Fe ratios, as is the case with hypernovae (Figure
\ref{fig:vssolarhn}) as pointed out by Umeda, Nomoto \& Nakamura
(2000).  Another feature of CS22873-139 is enhanced Ti/Fe ([Ti/Fe]
$\sim + 0.6$; Nordstr\"om et al. 2000; Spite et al. 2000), which could
be explained by a hypernova explosion.

It has been pointed out that Ti is deficient in Galactic chemical
evolution models using supernova yields currently available (e.g.,
Timmes et al. 1995; Thielemann et al. 1996), especially at [Fe/H]
$\lsim -1$ when SNe Ia have not contributed to the chemical evolution.  
However, if the contribution from hypernovae to
Galactic chemical evolution is relatively large (or supernovae are
more energetic than the typical value of 1 $\times$ $10^{51}$ erg),
this problem could be relaxed.  As we have seen in the previous
section, the $\alpha$-rich freezeout is enhanced in hypernovae because
nucleosynthesis proceeds under the circumstance of lower densities and
incomplete Si-burning occurs in more extended regions.  Thus,
$^{40}$Ca, $^{44}$Ca, and $^{48}$Ti are produced and could be ejected
into space more abundantly.

\subsection{Iron-Peak Elements}
\label{subsec:abundmp}

McWilliam et al. (1995) and Ryan et al. (1996) found a peculiar
abundance pattern in the iron-peak elements in metal-poor stars of
[Fe/H] $\lsim -2$. That is, [Cr/Fe] and [Mn/Fe] increase with
increasing [Fe/H], while [Co/Fe] shows the opposite trend and
decrease.  This trend cannot be explained with the conventional
chemical evolution model that uses previous nucleosynthesis yields
(e.g., Tsujimoto et al.  1995; Woosley \& Weaver 1995).

Nakamura et al. (1999) have shown that this trend of decreasing Cr and
Mn with increasing Co is reproduced by decreasing the mass cut
between the ejecta and the collapsed star for the same explosion
model. This is because Co is mostly produced in complete Si-burning
regions, while Mn and Cr are mainly produced in the outer incomplete
Si-burning region.  If the mass cut is located at smaller $M_r$, 
the mass ratio between the complete and incomplete Si-burning region
is larger.  Therefore, mass cuts at smaller $M_r$
increase the Co fraction but decrease the Mn and Cr fractions in the
ejecta.  Nakamura et al. (1999) have also shown that the observed
trend with respect to [Fe/H] may be explained if the mass cut tends to
be smaller in $M_r$ for the larger mass progenitor.

Here, we investigate whether the observed trend of these iron-peak
elements in metal-poor stars can be explained with the the abundance
pattern of hypernovae.  In Table \ref{tab:arip}, we summarize the
abundance ratios of iron-peak elements in the ejecta.  In Figure
\ref{fig:comn}, we plot [Mn/Fe] vs. [Co/Fe] and [Cr/Fe] for the 16
$M_\odot$ He star models with $E = (1$ - $30)$ $\times$ $ 10^{51}$ ergs,
and 8 $M_\odot$ He star models with $E = (1 $ - $10)$ $\times$ $
10^{51}$ ergs. This figure clearly shows the correlation between
[Mn/Fe] and [Cr/Fe], and anti-correlation between [Mn/Fe] and [Co/Fe],
which are the same trends as observed in the metal-poor stars.

To understand the dependence on $E$, let us compare the 
models with 1 $ \times$ $ 10^{51}$ ergs
and 10 $ \times$ $ 10^{51}$ ergs which have almost the same mass cut (Table
\ref{tab:masscuthn}).  In the model with $E =$ 10 $ \times$ $ 10^{51}$
ergs, both complete and incomplete Si-burning regions shift
outward in mass compared with $E =$ 1 $ \times$ $ 10^{51}$
ergs because of a larger explosion energy.  Thus, the model
with larger $E$ has a larger mass ratio between the
complete and incomplete Si-burning regions.  This relation between 
$E$ and the mass ratio does not
hold if the explosion energy is too large. For example, in the 16
$M_\odot$ He star models with 100 $ \times$ $ 10^{51}$ ergs and 8
$M_\odot$ He star models with more than 30 $ \times$ $ 10^{51}$ ergs,
incomplete Si-burning extends so far out that Mn and Cr increase too
much to fit the metal-poor star data. For this reason, we do not 
include these models in Figure \ref{fig:comn}.

In metal-poor stars, [Mn/Fe] increases with [Fe/H].  Hypernova yields
are consistent with this trend if hypernovae with larger $E$ induce
the formation of stars with smaller [Fe/H]. This supposition is
reasonable because the mass of interstellar hydrogen gathered by a
hypernova is roughly proportional to $E$ (Cioffi et al. 1988;
Shigeyama \& Tsujimoto 1998) and the ratio of the ejected iron mass to
$E$ is smaller for hypernovae than for canonical supernovae.

The amounts of iron-peak elements and their ratios in the ejecta of
hypernovae depend on the mass cut that is still uncertain.
How the abundance ratios depend on the mass cut is discussed in 
Umeda \& Nomoto (2001).  If we adopt a smaller
mass cut, a larger amount of $^{56}$Ni is ejected and
the abundance ratios can be different from Table
\ref{tab:arip}.  To reduce the uncertainty in the prediction,
it is necessary to observe more hypernovae.  We can determine the
amount of $^{56}$Ni ejected from a hypernova by the light curve and
spectral fitting (e.g., Iwamoto et al. 2000).  Combining this
information with nucleosynthesis calculations, we can estimate the
mass cut as in the case of SN1998bw (Nakamura et al. 2001).

\subsection{Abundances in Starburst Galaxies}

X-ray emissions from the starburst galaxy M82 were observed with ASCA
and the abundances of several heavy elements were obtained (Tsuru et
al. 1997).  Tsuru et al. (1997) found that the overall metallicity of
M82 is quite low, i.e., O/H and Fe/H are only 0.06 - 0.05 times solar,
while Si/H and S/H are $\sim$ 0.40 - 0.47 times solar.  This implies
that the abundance ratios are peculiar, i.e., the ratio O/Fe is about
solar, while the ratios of Si and S relative to O and Fe are as high
as $\sim$ 6 - 8.  These ratios are very different from those ratios in
SNe II.  The age of M82 is estimated to be $\lsim 10^8$ years, which
is too young for Type Ia supernovae to contribute to enhance Fe
relative to O.  Tsuru et al. (1997) also estimated that the explosion
energy required to produce the observed amount of hot plasma per
oxygen mass is significantly larger than that of normal SNe II (here
the oxygen mass dominates the total mass of the heavy elements).
Tsuru et al. (1997) thus concluded that neither SN Ia nor SN II can
reproduce the observed abundance pattern of M82.

Compared with normal SNe II, the important characteristic of hypernova
nucleosynthesis is the large Si/O, S/O, and Fe/O ratios.
Quantitatively, the enhancement of Si and S over O in our current
hypernova models are smaller than those observed in M82, but possible
aspherical effects might enhance the Si/O and S/O ratios over the
spherical models.  Also, the larger Si/O and S/O ratios are
yielded, if the progenitors of hypernovae have smaller
metallicities (Nakamura et al. 2000).
Hypernovae could also produce larger $E$ per oxygen
mass than normal SNe II.  We therefore suggest that hypernova
explosions may make important contributions to the metal enrichment
and energy input to the interstellar matter in M82.  If the IMF of the
star burst is relatively flat compared with Salpeter IMF, the
contribution of very massive stars and thus hypernovae could be much
larger than in our Galaxy.

\subsection{Abundances in a Black Hole Binary}

X-ray Nova Sco (GRO J1655-40), which consists of a massive black hole
and a low mass companion (e.g., Brandt et al. 1995; Nelemans et
al. 2000), also exhibits what could be the nucleosynthesis signature
of a hypernova explosion.
The companion star is enriched with Ti, S, Si, Mg, and O but not much
Fe (Israelian et al. 1999).  This is compatible with heavy element
ejection from a black hole progenitor.  In order to eject large amount
of Ti, S, and Si and to have at least $\sim$ 4 $M_\odot$ below mass
cut and thus form a massive black hole, the explosion
would need to be highly
energetic (Figure \ref{fig:donehn}; Israelian et al. 1999; Brown et
al. 2000; Podsiadlowski et al. 2000).  A hypernova explosion with
the mass cut at large $M_r$ ejects a relatively small mass Fe and would be
consistent with these observed abundance features.
Alternatively, if the explosion which resulted
from the formation of the black hole in Nova Sco was asymmetric,
then it is likely that the companion star captured
material ejected in the direction away from the strong shock which
contained relatively little Fe compared with the ejecta in the strong
shock (Maeda et al. 2000).

\section{Summary}\label{sec:summary}

We investigated explosive nucleosynthesis in hypernovae,
that is, hyper-energetic supernovae.
Detailed nucleosynthesis calculations were performed and compared to those of
ordinary core-collapse supernovae.  We also studied implications to
Galactic chemical evolution and the abundances in metal-poor stars.

We demonstrated, using the SN1998bw data, that progenitor of
hypernovae must be massive based on the information from light curve
and nucleosynthesis calculations of $^{56}$Ni, without using spectral
information. This argument is useful for cases where reliable spectral
information is not available.

In hypernovae, both complete and incomplete Si-burning takes place in
more extended, and hence, lower density regions, so that the
alpha-rich freezeout is enhanced in comparison with normal supernova
nucleosynthesis.  Thus $^{44}$Ca, $^{48}$Ti, and $^{64}$Zn are produced more
abundantly than in canonical supernovae.  Oxygen and carbon burning
also takes place in more extended regions for the larger explosion
energy, hence the yield of these fuel elements are smaller than
canonical supernovae.

We found that hypernova nucleosynthetic yields are compatible with the
observed Cr, Mn, and Fe abundances of metal-poor halo stars, provided
that the explosion energy is $\sim$ 10$^{52}$ ergs. Our results imply
that heavier progenitors allow larger explosion energy to match the
abundance pattern of metal-poor stars, although further investigations
are necessary for firm conclusion. Our results suggest that some
metal-poor halo stars may bear the abundance pattern of a single
hypernova.

We also suggest that the problem of the deficit of Ti in the Galactic
chemical evolution with the supernova yields currently available could
be settled by the contribution of hypernovae, in which $\alpha$-rich
freezeout is enhanced and large amounts of $\alpha$-elements are
ejected.  The abundance pattern of the starburst galaxy M82,
characterized by abundant Si and S relative to O and Fe, may be
attributed to hypernova explosions if the IMF is relatively flat, and
thus the contribution of massive stars to the galactic chemical
evolution is large.  The black hole binary X-ray Nova Sco (GRO
J1655-40) may also be affected by a hypernova explosion, which can
eject large amount of Ti, S, and Si and leave a massive black hole 
as a compact remnant.

In this paper, we performed calculations assuming spherical symmetry
and thus did not take account of non-spherical effects.  For example,
if a jet is associated with hypernovae, the explosive shock may be
pointed and stronger in this direction.  In this case, the high
explosion energies along the jet should have the properties of
nucleosynthesis in hypernovae discussed above and orthogonal to the
jet may be more like ordinary supernovae.
In such conditions, a hypernova could make ejecta which have a range
of abundance ratios (as seen in Figure 6) in different directions.
If those ejecta in various directions interact with interstellar
matter and form mixed materials of various mass ratios, stars which
could form from such interactions would have a range of metallicity
and abundance ratios among iron-peak elements.  For example, the
ejecta in the jet direction has larger amount of complete Si-burning
products (i.e., larger [Co/Fe]) and are mixed with larger amount of
hydrogen (i.e., producing smaller [Fe/H]) because of higher velocities
than in other directions.  Then such explosions might be responsible
for the abundance trends in the metal-poor stars.  Multi-dimensional
simulations are needed to investigate these possibilities.

\bigskip

We would like to thank Drs. B. Nordstr\"om, F. Spite, T. Tsuru,
G. Israelian, G. Brown, Ph. Podsiadlowski, P. Mazzali, and B. Schmidt
for stimulating discussion on the observational indication of
hypernova nucleosynthesis.
This work has been supported in part by the Grant-in-Aid for
Scientific Research (12640233, 12740122) and COE research (07CE2002)
of the Japanese Ministry of Education, Science, Culture, and Sports,
Swiss National Science Foundation grant 2000-53798.98, U.S. National
Aeronautics and Space Administration grant NAG5-8405 and the Joint
Institute for Heavy Ion Research, which has as member institutions the
University of Tennessee, Vanderbilt University, and the Oak Ridge
National Laboratory.  ORNL is managed by UT-Battelle, LLC, for the
U.S. Department of Energy under contract DE-AC05-00OR22725.

\clearpage

\begin{figure}
\epsscale{.6}
\plotone{fig2a.epsi}
\vspace{1cm}
\plotone{fig2c.epsi}
\caption{
Nucleosynthesis in hypernovae and normal core-collapse supernovae
for explosion energies of $E =$  100 (top left), 30 (top right),
10 (bottom left), and 1 (bottom right) $ \times$ $ 10^{51}$ ergs.
The progenitor model is the 16$M_{\odot}$ He core model.
\label{fig:donehn}}
\end{figure}

\addtocounter{figure}{-1}
\begin{figure}
\epsscale{.6}
\plotone{fig2b.epsi}
\vspace{1cm}
\plotone{fig2d.epsi}
\caption{
Nucleosynthesis in hypernovae and normal core-collapse supernovae
for explosion energies of $E =$
 100 (top left), 30 (top right),
10 (bottom left), and 1 (bottom right) $ \times$ $ 10^{51}$ ergs.
The progenitor model is the 16$M_{\odot}$ He core model.
\label{fig:donehn2}}
\end{figure}

\begin{figure}
\epsscale{.7}
\plotone{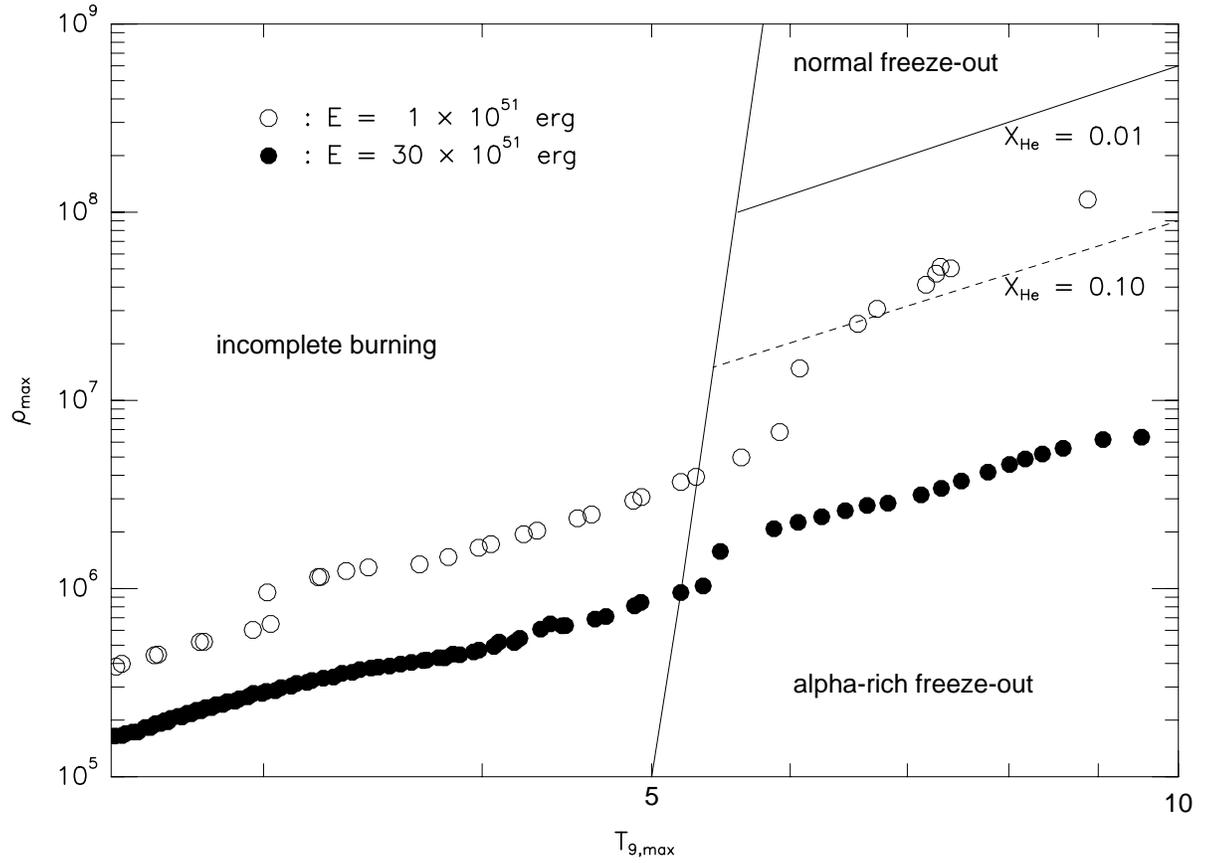}
\caption{
The $\rho$ - $T$ conditions of individual mass zones at their
temperature maximum in hypernovae ($E=$ 30 $ \times$ $ 10^{51}$ ergs:
filled circles) and normal supernovae ($E=$ 1 $ \times$ $ 10^{51}$erg:
open circles).  The lines that separate the Si-burning regimes and
contour lines for constant $^4$He mass fractions are taken from
Thielemann et al. (1998).
\label{fig:maxrt}}
\end{figure}

\begin{figure}
\epsscale{.6}
\plotone{fig4a.epsi}
\vspace{1cm}
\epsscale{.6}
\plotone{fig4c.epsi}
\caption{
Abundances of stable isotopes relative to the solar values
for $E =$ 100 (top left), 30 (top right),
10 (bottom left), and 1 (bottom right) $ \times$ $ 10^{51}$ ergs.
The progenitor model is the 16$M_{\odot}$ He core model
(H-rich envelope is not included).
\label{fig:vssolarhn}}
\end{figure}

\addtocounter{figure}{-1}
\begin{figure}
\epsscale{.6}
\plotone{fig4b.epsi}
\vspace{1cm}
\epsscale{.6}
\plotone{fig4d.epsi}
\caption{
Abundances of stable isotopes relative to the solar values
for $E =$ 100 (top left), 30 (top right),
10 (bottom left), and 1 (bottom right) $ \times$ $ 10^{51}$ ergs.
The progenitor model is the 16$M_{\odot}$ He core model
(H-rich envelope is not included).
\label{fig:vssolarhn2}}
\end{figure}

\begin{figure}
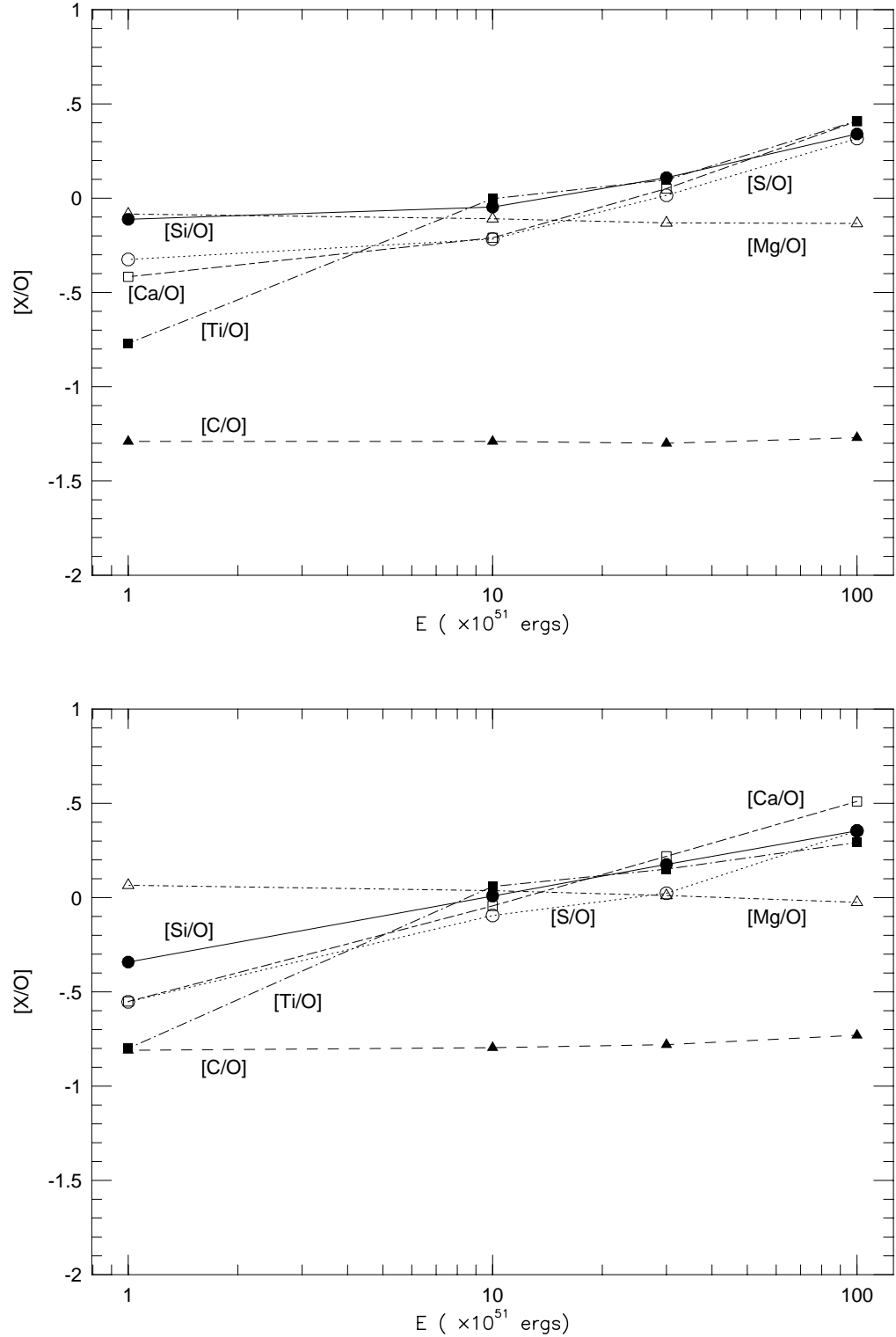

\epsscale{.6}
\plotone{fig5a.epsi}
\vspace{1cm}
\epsscale{.6}
\plotone{fig5b.epsi}
\caption{
Abundance ratios to oxygen relative to solar values
as a function of the explosion energy $E$.
The progenitor model is the 16$M_{\odot}$ He core model (above)
and the 8$M_{\odot}$ He core model (below).
Products from H-rich envelope are not included in both models.
\label{fig:ratiohn}}
\end{figure}

\begin{figure}
\epsscale{.6}
\plotone{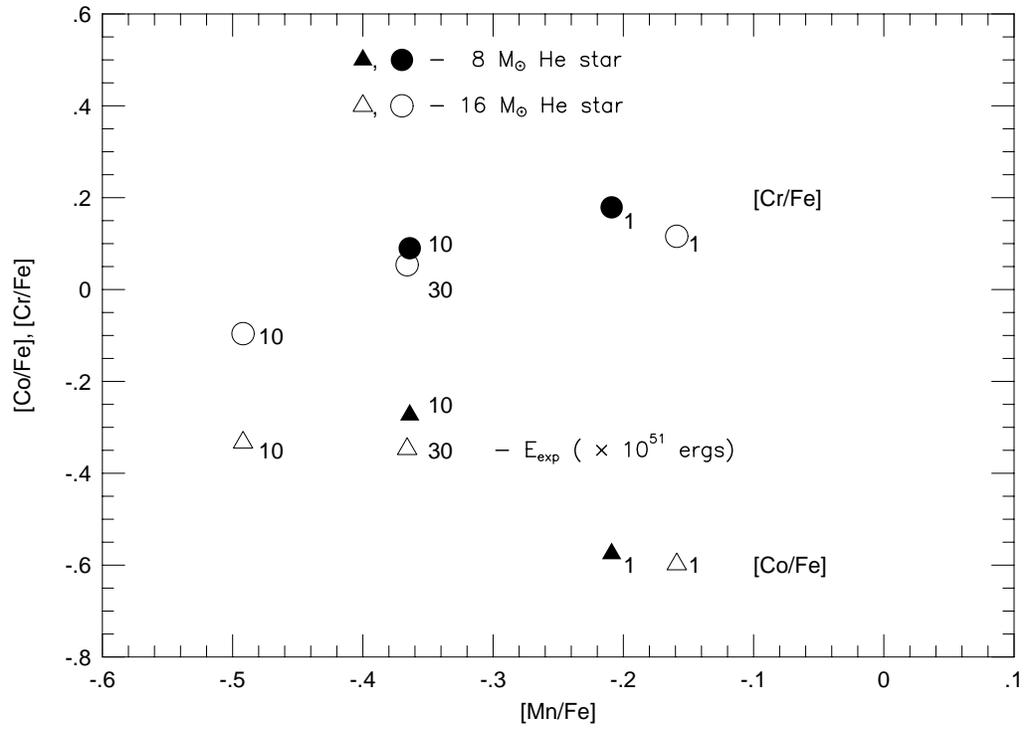}
\caption{[Mn/Fe] vs. [Co/Fe] and [Mn/Fe] for the
16 $M_\odot$ He star models with $E = (1.0$ - $30)
\times$ $ 10^{51}$ ergs, and 8 $M_\odot$ He star models with
$E = (1.0$ - $10) \times$ $10^{51}$ ergs.
\label{fig:comn}}
\end{figure}

\clearpage

\begin{deluxetable}{crrrr}
\tablecaption{Mass cuts chosen in this paper.
\label{tab:masscuthn}}
\tablewidth{0pt}
\tablehead{
\colhead{{$E$ ($\times$ $ 10^{51}$ ergs)}}
&\colhead{100}&\colhead{30}&\colhead{10}&\colhead{1} \\
\colhead{Progenitor}
}
\startdata
16$M_{\odot}$ He core & 3.5$M_{\odot}$ & 2.9$M_{\odot}$ &
2.3$M_{\odot}$ & 2.3$M_{\odot}$ \\
10$M_{\odot}$ He core & 2.5$M_{\odot}$ & 2.5$M_{\odot}$ &
1.7$M_{\odot}$ & 1.7$M_{\odot}$ \\
8$M_{\odot}$ He core & 2.4$M_{\odot}$ & 1.9$M_{\odot}$ &
1.7$M_{\odot}$ & 1.7$M_{\odot}$ \\
6$M_{\odot}$ He core & 2.3$M_{\odot}$ & 1.8$M_{\odot}$ &
1.6$M_{\odot}$ & 1.6$M_{\odot}$ \\
\enddata
\end{deluxetable}

\begin{deluxetable}{crrrrrcrrrr}
\scriptsize
\tablecaption{Nucleosynthesis products ($M_\odot$)
of hypernovae (and normal supernovae)
after decay of radioactive species for various explosion energies.
The progenitor model is the 16$M_{\odot}$ He core model.
\label{doneisohn16a}}
\tablewidth{0pt}
\tablehead{
\colhead{{$E$ ($\times$ $ 10^{51}$ ergs)}}
&\colhead{100}&\colhead{30}&\colhead{10}&\colhead{1}&&
\colhead{$E$}
&\colhead{100}&\colhead{30}&\colhead{10}&\colhead{1}\\
\colhead{species}&&&&&&\colhead{species}}
\startdata
$^{ 4}$He & 2.65     & 2.20     & 2.11     & 1.96     & &
$^{43}$Ca & 1.89E-05 & 1.25E-05 & 1.27E-05 & 1.46E-06 \\
$^{12}$C  & 1.05E-01 & 1.25E-01 & 1.38E-01 & 1.49E-01 & &
$^{44}$Ca & 3.24E-03 & 1.85E-03 & 1.89E-03 & 1.53E-04 \\
$^{13}$C  & 5.97E-07 & 2.75E-07 & 2.35E-07 & 1.56E-08 & &
$^{46}$Ca & 6.19E-10 & 5.04E-10 & 3.18E-10 & 1.20E-10 \\
$^{14}$N  & 1.22E-04 & 1.22E-04 & 1.22E-04 & 1.21E-04 & &
$^{48}$Ca & 7.03E-13 & 2.17E-13 & 2.45E-13 & 2.35E-14 \\
$^{15}$N  & 9.23E-08 & 2.44E-08 & 1.83E-08 & 1.01E-08 & &
$^{45}$Sc & 1.59E-06 & 9.59E-07 & 7.94E-07 & 5.14E-07 \\
$^{16}$O  & 6.21     & 7.77     & 8.49     & 9.08     & &
$^{46}$Ti & 1.39E-04 & 8.24E-05 & 5.90E-05 & 7.45E-05 \\
$^{17}$O  & 2.04E-07 & 2.30E-07 & 1.85E-07 & 8.26E-08 & &
$^{47}$Ti & 9.25E-05 & 6.44E-05 & 5.97E-05 & 5.50E-06 \\
$^{18}$O  & 2.37E-07 & 1.71E-07 & 1.58E-07 & 1.63E-07 & &
$^{48}$Ti & 4.51E-03 & 2.74E-03 & 2.40E-03 & 3.65E-04 \\
$^{19}$F  & 4.32E-08 & 4.12E-09 & 6.08E-10 & 4.31E-10 & &
$^{49}$Ti & 7.45E-05 & 4.20E-05 & 2.61E-05 & 2.08E-05 \\
$^{20}$Ne & 2.97E-01 & 4.44E-01 & 5.53E-01 & 6.58E-01 & &
$^{50}$Ti & 4.88E-09 & 2.94E-09 & 1.76E-09 & 2.12E-10 \\
$^{21}$Ne & 1.62E-03 & 2.12E-03 & 2.09E-03 & 2.34E-03 & &
$^{50}$V  & 1.24E-08 & 7.26E-09 & 5.12E-09 & 3.85E-09 \\
$^{22}$Ne & 3.78E-02 & 5.23E-02 & 5.69E-02 & 5.74E-02 & &
$^{51}$V  & 1.54E-04 & 9.75E-05 & 6.88E-05 & 4.94E-05 \\
$^{23}$Na & 9.05E-03 & 1.55E-02 & 2.03E-02 & 2.37E-02 & &
$^{50}$Cr & 6.78E-04 & 4.41E-04 & 3.20E-04 & 3.66E-04 \\
$^{24}$Mg & 2.26E-01 & 2.83E-01 & 3.22E-01 & 3.59E-01 & &
$^{52}$Cr & 1.76E-02 & 1.07E-02 & 7.30E-03 & 4.07E-03 \\
$^{25}$Mg & 2.73E-02 & 3.51E-02 & 4.10E-02 & 4.84E-02 & &
$^{53}$Cr & 1.67E-03 & 9.34E-04 & 6.24E-04 & 4.80E-04 \\
$^{26}$Mg & 5.97E-02 & 7.60E-02 & 9.07E-02 & 1.07E-01 & &
$^{54}$Cr & 3.83E-07 & 2.28E-07 & 1.61E-07 & 1.85E-07 \\
$^{27}$Al & 3.85E-02 & 5.78E-02 & 6.92E-02 & 7.97E-02 & &
$^{55}$Mn & 6.05E-03 & 3.42E-03 & 2.48E-03 & 1.95E-03 \\
$^{28}$Si & 9.11E-01 & 6.37E-01 & 4.63E-01 & 4.21E-01 & &
$^{54}$Fe & 6.92E-02 & 4.13E-02 & 2.94E-02 & 2.71E-02 \\
$^{29}$Si & 3.84E-02 & 4.57E-02 & 4.98E-02 & 5.42E-02 & &
$^{56}$Fe & 8.47E-01 & 6.93E-01 & 6.79E-01 & 2.35E-01 \\
$^{30}$Si & 5.59E-02 & 5.52E-02 & 4.99E-02 & 4.38E-02 & &
$^{57}$Fe & 3.66E-02 & 2.87E-02 & 2.87E-02 & 7.10E-03 \\
$^{31}$P  & 6.55E-03 & 6.90E-03 & 6.56E-03 & 6.05E-03 & &
$^{58}$Fe & 1.34E-07 & 7.62E-08 & 5.26E-08 & 5.13E-08 \\
$^{32}$S  & 5.30E-01 & 3.29E-01 & 2.09E-01 & 1.76E-01 & &
$^{59}$Co & 1.59E-03 & 9.05E-04 & 9.04E-04 & 1.79E-04 \\
$^{33}$S  & 1.63E-03 & 1.28E-03 & 9.20E-04 & 7.48E-04 & &
$^{58}$Ni & 5.78E-02 & 4.16E-02 & 4.21E-02 & 1.22E-02 \\
$^{34}$S  & 2.65E-02 & 1.96E-02 & 1.37E-02 & 1.03E-02 & &
$^{60}$Ni & 3.09E-02 & 1.93E-02 & 2.27E-02 & 4.21E-03 \\
$^{36}$S  & 2.33E-05 & 1.89E-05 & 1.37E-05 & 7.86E-06 & &
$^{61}$Ni & 1.03E-03 & 1.06E-03 & 1.40E-03 & 2.22E-04 \\
$^{35}$Cl & 5.87E-04 & 4.54E-04 & 3.70E-04 & 3.93E-04 & &
$^{62}$Ni & 1.06E-02 & 8.56E-03 & 9.72E-03 & 1.74E-03 \\
$^{37}$Cl & 1.63E-04 & 1.08E-04 & 8.12E-05 & 1.08E-04 & &
$^{64}$Ni & 1.48E-11 & 3.50E-12 & 4.04E-12 & 6.10E-13 \\
$^{36}$Ar & 1.02E-01 & 5.90E-02 & 3.58E-02 & 2.83E-02 & &
$^{63}$Cu & 2.09E-05 & 1.53E-05 & 1.53E-05 & 1.97E-06 \\
$^{38}$Ar & 1.17E-02 & 7.56E-03 & 5.43E-03 & 7.85E-03 & &
$^{65}$Cu & 1.86E-05 & 1.14E-05 & 1.24E-05 & 1.36E-06 \\
$^{40}$Ar & 3.21E-07 & 2.71E-07 & 1.84E-07 & 9.94E-08 & &
$^{64}$Zn & 1.45E-04 & 9.87E-05 & 1.19E-04 & 1.95E-05 \\
$^{39}$K  & 6.08E-04 & 3.98E-04 & 3.12E-04 & 3.37E-04 & &
$^{66}$Zn & 1.47E-04 & 1.62E-04 & 2.00E-04 & 3.28E-05 \\
$^{41}$K  & 4.26E-05 & 2.71E-05 & 2.14E-05 & 2.42E-05 & &
$^{67}$Zn & 9.06E-07 & 6.92E-07 & 6.15E-07 & 4.89E-08 \\
$^{40}$Ca & 9.87E-02 & 5.42E-02 & 3.16E-02 & 2.20E-02 & &
$^{68}$Zn & 2.40E-07 & 1.30E-07 & 1.15E-07 & 1.47E-08 \\
$^{42}$Ca & 3.49E-04 & 2.13E-04 & 1.55E-04 & 2.31E-04 & &
$^{70}$Zn & 5.89E-12 & 2.55E-12 & 4.60E-12 & 4.61E-13 \\
\enddata
\end{deluxetable}

\begin{deluxetable}{crrrrrcrrrr}
\scriptsize
\tablecaption{
Nucleosynthesis products ($M_\odot$)
of hypernovae (and normal supernovae)
after decay of radioactive species for various explosion energies.
The progenitor model is the 10$M_{\odot}$ He core model.
\label{doneisohn10a}}
\tablewidth{0pt}
\tablehead{
\colhead{{$E$ ($\times$ $ 10^{51}$ ergs)}}
&\colhead{100}&\colhead{30}&\colhead{10}&\colhead{1}&&
\colhead{$E$}
&\colhead{100}&\colhead{30}&\colhead{10}&\colhead{1}\\
\colhead{species}&&&&&&\colhead{species}}
\startdata
$^{ 4}$He & 3.16     & 2.45     & 2.48     & 2.35     & &
$^{43}$Ca & 2.35E-05 & 5.02E-06 & 8.67E-06 & 2.10E-06 \\
$^{12}$C  & 2.01E-01 & 2.30E-01 & 2.43E-01 & 2.55E-01 & &
$^{44}$Ca & 3.16E-03 & 6.45E-04 & 1.22E-03 & 1.81E-04 \\
$^{13}$C  & 2.03E-05 & 8.34E-06 & 1.66E-06 & 3.55E-07 & &
$^{46}$Ca & 3.25E-10 & 2.43E-10 & 1.99E-10 & 7.41E-11 \\
$^{14}$N  & 6.46E-04 & 6.38E-04 & 6.38E-04 & 6.38E-04 & &
$^{48}$Ca & 3.69E-12 & 1.04E-13 & 1.17E-13 & 1.26E-14 \\
$^{15}$N  & 1.61E-07 & 6.14E-08 & 3.80E-08 & 1.87E-08 & &
$^{45}$Sc & 1.00E-06 & 6.83E-07 & 9.60E-07 & 6.71E-07 \\
$^{16}$O  & 1.92     & 2.88     & 3.46     & 3.96     & &
$^{46}$Ti & 8.33E-05 & 6.48E-05 & 4.31E-05 & 1.45E-04 \\
$^{17}$O  & 9.30E-07 & 3.78E-07 & 2.82E-07 & 1.25E-07 & &
$^{47}$Ti & 1.09E-04 & 2.21E-05 & 6.89E-05 & 1.19E-05 \\
$^{18}$O  & 3.49E-05 & 5.00E-05 & 6.33E-05 & 7.73E-05 & &
$^{48}$Ti & 4.27E-03 & 1.22E-03 & 1.90E-03 & 2.92E-04 \\
$^{19}$F  & 3.55E-08 & 4.41E-08 & 1.11E-08 & 4.70E-09 & &
$^{49}$Ti & 4.73E-05 & 3.86E-05 & 3.08E-05 & 1.53E-05 \\
$^{20}$Ne & 1.62E-02 & 1.43E-01 & 2.82E-01 & 4.59E-01 & &
$^{50}$Ti & 2.31E-09 & 1.97E-09 & 1.22E-09 & 9.58E-10 \\
$^{21}$Ne & 3.99E-04 & 6.82E-04 & 1.31E-03 & 1.88E-03 & &
$^{50}$V  & 6.57E-09 & 5.29E-09 & 3.34E-09 & 1.32E-08 \\
$^{22}$Ne & 3.56E-02 & 4.99E-02 & 5.90E-02 & 6.33E-02 & &
$^{51}$V  & 1.36E-04 & 7.57E-05 & 8.47E-05 & 6.21E-05 \\
$^{23}$Na & 1.97E-03 & 4.65E-03 & 1.09E-02 & 1.79E-02 & &
$^{50}$Cr & 4.21E-04 & 3.81E-04 & 3.13E-04 & 7.97E-04 \\
$^{24}$Mg & 6.24E-02 & 1.23E-01 & 1.62E-01 & 2.06E-01 & &
$^{52}$Cr & 1.29E-02 & 8.50E-03 & 8.00E-03 & 2.81E-03 \\
$^{25}$Mg & 4.01E-03 & 1.18E-02 & 1.94E-02 & 2.89E-02 & &
$^{53}$Cr & 1.09E-03 & 8.89E-04 & 7.40E-04 & 5.32E-04 \\
$^{26}$Mg & 2.06E-02 & 2.88E-02 & 3.13E-02 & 4.21E-02 & &
$^{54}$Cr & 2.46E-07 & 1.95E-07 & 1.34E-07 & 9.90E-07 \\
$^{27}$Al & 6.30E-03 & 1.58E-02 & 2.48E-02 & 3.51E-02 & &
$^{55}$Mn & 4.14E-03 & 3.27E-03 & 2.90E-03 & 2.26E-03 \\
$^{28}$Si & 5.66E-01 & 5.08E-01 & 4.01E-01 & 3.34E-01 & &
$^{54}$Fe & 4.36E-02 & 3.62E-02 & 3.23E-02 & 4.32E-02 \\
$^{29}$Si & 1.18E-02 & 1.55E-02 & 1.73E-02 & 1.92E-02 & &
$^{56}$Fe & 7.15E-01 & 4.66E-01 & 5.64E-01 & 2.00E-01 \\
$^{30}$Si & 2.79E-02 & 2.62E-02 & 2.44E-02 & 1.94E-02 & &
$^{57}$Fe & 3.56E-02 & 1.70E-02 & 2.93E-02 & 8.91E-03 \\
$^{31}$P  & 3.09E-03 & 3.37E-03 & 3.42E-03 & 3.06E-03 & &
$^{58}$Fe & 8.83E-08 & 6.36E-08 & 4.38E-08 & 2.06E-07 \\
$^{32}$S  & 3.35E-01 & 2.70E-01 & 2.00E-01 & 1.32E-01 & &
$^{59}$Co & 1.99E-03 & 4.38E-04 & 1.22E-03 & 3.32E-04 \\
$^{33}$S  & 1.22E-03 & 1.09E-03 & 8.90E-04 & 8.44E-04 & &
$^{58}$Ni & 6.07E-02 & 2.00E-02 & 9.89E-02 & 2.99E-02 \\
$^{34}$S  & 1.87E-02 & 1.51E-02 & 1.14E-02 & 1.34E-02 & &
$^{60}$Ni & 3.54E-02 & 1.07E-02 & 1.58E-02 & 5.02E-03 \\
$^{36}$S  & 4.44E-05 & 9.08E-06 & 7.78E-06 & 5.14E-06 & &
$^{61}$Ni & 9.76E-04 & 4.50E-04 & 1.33E-03 & 3.79E-04 \\
$^{35}$Cl & 7.29E-04 & 3.81E-04 & 3.24E-04 & 4.66E-04 & &
$^{62}$Ni & 1.12E-02 & 3.41E-03 & 1.87E-02 & 5.56E-03 \\
$^{37}$Cl & 1.25E-04 & 9.04E-05 & 6.58E-05 & 1.31E-04 & &
$^{64}$Ni & 4.57E-11 & 2.68E-12 & 2.76E-12 & 4.15E-13 \\
$^{36}$Ar & 6.44E-02 & 4.84E-02 & 3.50E-02 & 1.90E-02 & &
$^{63}$Cu & 2.26E-05 & 4.57E-06 & 3.57E-05 & 6.90E-06 \\
$^{38}$Ar & 6.83E-03 & 5.86E-03 & 3.96E-03 & 1.36E-02 & &
$^{65}$Cu & 2.14E-05 & 3.65E-06 & 1.40E-05 & 2.98E-06 \\
$^{40}$Ar & 2.36E-06 & 1.26E-07 & 1.07E-07 & 6.92E-08 & &
$^{64}$Zn & 1.60E-04 & 5.27E-05 & 6.99E-05 & 2.36E-05 \\
$^{39}$K  & 4.10E-04 & 2.87E-04 & 2.48E-04 & 4.94E-04 & &
$^{66}$Zn & 1.48E-04 & 5.48E-05 & 3.59E-04 & 1.21E-04 \\
$^{41}$K  & 2.80E-05 & 2.05E-05 & 1.85E-05 & 3.07E-05 & &
$^{67}$Zn & 9.95E-07 & 1.51E-07 & 1.56E-06 & 2.72E-07 \\
$^{40}$Ca & 6.08E-02 & 4.43E-02 & 3.20E-02 & 1.36E-02 & &
$^{68}$Zn & 3.01E-07 & 4.01E-08 & 4.61E-07 & 1.20E-07 \\
$^{42}$Ca & 2.01E-04 & 1.59E-04 & 1.10E-04 & 4.46E-04 & &
$^{70}$Zn & 1.79E-11 & 2.16E-12 & 1.17E-12 & 2.14E-13 \\
\enddata
\end{deluxetable}

\begin{deluxetable}{crrrrrcrrrr}
\scriptsize
\tablecaption{
Nucleosynthesis products ($M_\odot$)
of hypernovae (and normal supernovae)
after decay of radioactive species for various explosion energies.
The progenitor model is the 8$M_{\odot}$ He core model.
\label{doneisohn8a}}
\tablewidth{0pt}
\tablehead{
\colhead{{$E$ ($\times$ $ 10^{51}$ ergs)}}
&\colhead{100}&\colhead{30}&\colhead{10}&\colhead{1}&&
\colhead{$E$}
&\colhead{100}&\colhead{30}&\colhead{10}&\colhead{1}\\
\colhead{species}&&&&&&\colhead{species}}
\startdata
$^{ 4}$He & 2.25     & 2.21     & 2.07     & 1.95     & &
$^{43}$Ca & 2.96E-06 & 5.30E-06 & 6.49E-06 & 7.12E-07 \\
$^{12}$C  & 1.19E-01 & 1.32E-01 & 1.40E-01 & 1.49E-01 & &
$^{44}$Ca & 9.82E-04 & 7.35E-04 & 6.34E-04 & 6.70E-05 \\
$^{13}$C  & 7.54E-06 & 5.69E-06 & 4.06E-06 & 3.11E-09 & &
$^{46}$Ca & 7.66E-11 & 6.28E-11 & 4.47E-11 & 1.30E-11 \\
$^{14}$N  & 9.46E-04 & 9.46E-04 & 9.46E-04 & 9.46E-04 & &
$^{48}$Ca & 9.36E-14 & 1.52E-13 & 1.07E-13 & 4.62E-15 \\
$^{15}$N  & 5.73E-07 & 5.27E-07 & 4.09E-07 & 2.74E-09 & &
$^{45}$Sc & 6.50E-07 & 4.39E-07 & 2.29E-07 & 8.38E-08 \\
$^{16}$O  & 1.99     & 2.47     & 2.74     & 3.00     & &
$^{46}$Ti & 4.29E-05 & 2.64E-05 & 1.64E-05 & 4.63E-06 \\
$^{17}$O  & 2.89E-07 & 1.29E-07 & 8.82E-08 & 2.64E-08 & &
$^{47}$Ti & 1.53E-05 & 2.32E-05 & 2.32E-05 & 2.95E-06 \\
$^{18}$O  & 3.97E-03 & 5.22E-03 & 5.87E-03 & 6.69E-03 & &
$^{48}$Ti & 8.67E-04 & 9.90E-04 & 8.90E-04 & 1.31E-04 \\
$^{19}$F  & 9.63E-06 & 9.79E-06 & 7.27E-06 & 3.78E-10 & &
$^{49}$Ti & 2.45E-05 & 2.04E-05 & 1.20E-05 & 5.06E-06 \\
$^{20}$Ne & 1.63E-01 & 3.28E-01 & 4.52E-01 & 6.12E-01 & &
$^{50}$Ti & 7.63E-10 & 6.15E-10 & 3.71E-10 & 1.01E-10 \\
$^{21}$Ne & 1.23E-03 & 2.08E-03 & 2.67E-03 & 3.23E-03 & &
$^{50}$V  & 3.02E-09 & 1.92E-09 & 1.20E-09 & 3.38E-10 \\
$^{22}$Ne & 2.62E-02 & 3.19E-02 & 3.34E-02 & 3.39E-02 & &
$^{51}$V  & 3.82E-05 & 3.88E-05 & 3.18E-05 & 9.31E-06 \\
$^{23}$Na & 5.20E-03 & 1.02E-02 & 1.37E-02 & 1.85E-02 & &
$^{50}$Cr & 1.50E-04 & 1.25E-04 & 8.50E-05 & 3.81E-05 \\
$^{24}$Mg & 1.02E-01 & 1.34E-01 & 1.52E-01 & 1.68E-01 & &
$^{52}$Cr & 4.84E-03 & 4.47E-03 & 3.34E-03 & 1.24E-03 \\
$^{25}$Mg & 1.28E-02 & 2.26E-02 & 3.00E-02 & 4.01E-02 & &
$^{53}$Cr & 5.96E-04 & 4.64E-04 & 2.80E-04 & 1.18E-04 \\
$^{26}$Mg & 1.47E-02 & 1.79E-02 & 2.36E-02 & 3.24E-02 & &
$^{54}$Cr & 8.51E-08 & 6.49E-08 & 4.07E-08 & 1.26E-08 \\
$^{27}$Al & 1.05E-02 & 1.46E-02 & 1.71E-02 & 1.95E-02 & &
$^{55}$Mn & 1.95E-03 & 1.51E-03 & 9.73E-04 & 4.28E-04 \\
$^{28}$Si & 3.12E-01 & 2.56E-01 & 1.89E-01 & 8.91E-02 & &
$^{54}$Fe & 1.67E-02 & 1.28E-02 & 8.65E-03 & 3.65E-03 \\
$^{29}$Si & 8.04E-03 & 7.93E-03 & 7.70E-03 & 6.61E-03 & &
$^{56}$Fe & 2.37E-01 & 1.69E-01 & 1.98E-01 & 6.09E-02 \\
$^{30}$Si & 1.31E-02 & 1.14E-02 & 9.33E-03 & 5.48E-03 & &
$^{57}$Fe & 9.03E-03 & 6.53E-03 & 8.99E-03 & 1.71E-03 \\
$^{31}$P  & 1.88E-03 & 1.66E-03 & 1.30E-03 & 6.41E-04 & &
$^{58}$Fe & 3.40E-08 & 2.50E-08 & 1.36E-08 & 4.09E-09 \\
$^{32}$S  & 1.86E-01 & 1.35E-01 & 9.03E-02 & 3.47E-02 & &
$^{59}$Co & 9.54E-04 & 5.04E-04 & 3.04E-04 & 4.65E-05 \\
$^{33}$S  & 8.19E-04 & 6.62E-04 & 4.76E-04 & 1.78E-04 & &
$^{58}$Ni & 1.92E-02 & 1.26E-02 & 7.78E-03 & 2.26E-03 \\
$^{34}$S  & 9.25E-03 & 7.22E-03 & 4.97E-03 & 1.76E-03 & &
$^{60}$Ni & 1.34E-02 & 9.16E-03 & 8.11E-03 & 1.21E-03 \\
$^{36}$S  & 2.74E-06 & 2.15E-06 & 1.62E-06 & 5.93E-07 & &
$^{61}$Ni & 1.43E-04 & 1.99E-04 & 3.15E-04 & 7.17E-05 \\
$^{35}$Cl & 2.80E-04 & 2.35E-04 & 1.73E-04 & 6.15E-05 & &
$^{62}$Ni & 1.79E-03 & 2.13E-03 & 2.32E-03 & 4.97E-04 \\
$^{37}$Cl & 6.09E-05 & 4.45E-05 & 2.93E-05 & 1.01E-05 & &
$^{64}$Ni & 4.21E-12 & 1.33E-10 & 2.84E-12 & 1.68E-13 \\
$^{36}$Ar & 3.86E-02 & 2.60E-02 & 1.63E-02 & 5.94E-03 & &
$^{63}$Cu & 1.45E-05 & 2.10E-05 & 4.61E-06 & 7.78E-07 \\
$^{38}$Ar & 3.42E-03 & 2.29E-03 & 1.52E-03 & 4.59E-04 & &
$^{65}$Cu & 1.53E-05 & 1.51E-05 & 3.77E-06 & 6.22E-07 \\
$^{40}$Ar & 4.25E-08 & 3.47E-08 & 2.49E-08 & 8.70E-09 & &
$^{64}$Zn & 3.42E-05 & 6.28E-05 & 6.01E-05 & 8.05E-06 \\
$^{39}$K  & 1.77E-04 & 1.39E-04 & 9.11E-05 & 2.60E-05 & &
$^{66}$Zn & 2.39E-05 & 4.44E-05 & 3.97E-05 & 1.20E-05 \\
$^{41}$K  & 1.46E-05 & 1.03E-05 & 6.21E-06 & 2.02E-06 & &
$^{67}$Zn & 1.06E-06 & 3.48E-06 & 1.57E-07 & 2.95E-08 \\
$^{40}$Ca & 4.07E-02 & 2.56E-02 & 1.53E-02 & 5.36E-03 & &
$^{68}$Zn & 9.53E-07 & 5.41E-06 & 4.26E-08 & 7.71E-09 \\
$^{42}$Ca & 1.03E-04 & 6.33E-05 & 4.01E-05 & 1.16E-05 & &
$^{70}$Zn & 1.31E-12 & 1.19E-12 & 2.51E-12 & 4.16E-14 \\
\enddata
\end{deluxetable}

\begin{deluxetable}{crrrrrcrrrr}
\scriptsize
\tablecaption{
Nucleosynthesis products ($M_\odot$)
of hypernovae (and normal supernovae)
after decay of radioactive species for various explosion energies.
The progenitor model is the 6$M_{\odot}$ He core model.
\label{doneisohn6a}}
\tablewidth{0pt}
\tablehead{
\colhead{{$E$ ($\times$ $ 10^{51}$ ergs)}}
&\colhead{100}&\colhead{30}&\colhead{10}&\colhead{1}&&
\colhead{$E$}
&\colhead{100}&\colhead{30}&\colhead{10}&\colhead{1}\\
\colhead{species}&&&&&&\colhead{species}}
\startdata
$^{ 4}$He & 2.17E+00 & 2.26E+00 & 2.24E+00 & 2.10E+00 & &
$^{43}$Ca & 7.02E-07 & 2.95E-06 & 7.40E-06 & 1.11E-06 \\
$^{12}$C  & 9.95E-02 & 1.06E-01 & 1.10E-01 & 1.14E-01 & &
$^{44}$Ca & 4.18E-05 & 3.27E-04 & 7.01E-04 & 8.81E-05 \\
$^{13}$C  & 1.37E-05 & 6.72E-06 & 6.46E-06 & 5.28E-10 & &
$^{46}$Ca & 3.10E-11 & 2.36E-11 & 1.61E-11 & 9.16E-12 \\
$^{14}$N  & 2.70E-03 & 2.71E-03 & 2.71E-03 & 2.71E-03 & &
$^{48}$Ca & 1.48E-13 & 1.42E-13 & 1.60E-13 & 1.12E-14 \\
$^{15}$N  & 5.23E-06 & 4.42E-06 & 3.43E-06 & 2.53E-10 & &
$^{45}$Sc & 6.26E-07 & 2.70E-07 & 2.35E-07 & 9.38E-08 \\
$^{16}$O  & 8.18E-01 & 1.10E+00 & 1.27E+00 & 1.48E+00 & &
$^{46}$Ti & 2.67E-05 & 1.76E-05 & 1.14E-05 & 3.87E-06 \\
$^{17}$O  & 9.46E-08 & 3.96E-08 & 2.71E-08 & 5.49E-09 & &
$^{47}$Ti & 2.00E-06 & 1.38E-05 & 3.07E-05 & 6.06E-06 \\
$^{18}$O  & 5.68E-03 & 6.89E-03 & 7.67E-03 & 8.68E-03 & &
$^{48}$Ti & 1.56E-04 & 4.90E-04 & 9.44E-04 & 1.55E-04 \\
$^{19}$F  & 1.60E-05 & 1.16E-05 & 8.99E-06 & 3.38E-11 & &
$^{49}$Ti & 1.47E-05 & 1.25E-05 & 8.82E-06 & 3.83E-06 \\
$^{20}$Ne & 3.69E-02 & 9.64E-02 & 1.45E-01 & 2.19E-01 & &
$^{50}$Ti & 3.66E-10 & 3.41E-10 & 2.34E-10 & 8.29E-11 \\
$^{21}$Ne & 6.85E-04 & 6.72E-04 & 5.98E-04 & 2.92E-04 & &
$^{50}$V  & 1.70E-09 & 1.10E-09 & 7.49E-10 & 2.65E-10 \\
$^{22}$Ne & 2.33E-02 & 2.71E-02 & 2.89E-02 & 2.93E-02 & &
$^{51}$V  & 1.53E-05 & 2.24E-05 & 3.51E-05 & 1.04E-05 \\
$^{23}$Na & 1.28E-03 & 1.41E-03 & 1.35E-03 & 1.10E-03 & &
$^{50}$Cr & 7.94E-05 & 6.92E-05 & 6.30E-05 & 3.39E-05 \\
$^{24}$Mg & 6.21E-02 & 9.60E-02 & 1.19E-01 & 1.48E-01 & &
$^{52}$Cr & 2.17E-03 & 2.36E-03 & 2.33E-03 & 9.43E-04 \\
$^{25}$Mg & 3.99E-03 & 8.45E-03 & 1.20E-02 & 1.78E-02 & &
$^{53}$Cr & 3.49E-04 & 2.78E-04 & 1.65E-04 & 8.44E-05 \\
$^{26}$Mg & 6.60E-03 & 9.09E-03 & 1.09E-02 & 1.65E-02 & &
$^{54}$Cr & 5.04E-08 & 4.33E-08 & 2.99E-08 & 1.19E-08 \\
$^{27}$Al & 5.67E-03 & 9.43E-03 & 1.21E-02 & 1.54E-02 & &
$^{55}$Mn & 1.18E-03 & 9.47E-04 & 5.87E-04 & 3.06E-04 \\
$^{28}$Si & 1.61E-01 & 1.56E-01 & 1.34E-01 & 8.94E-02 & &
$^{54}$Fe & 8.93E-03 & 7.64E-03 & 5.78E-03 & 3.01E-03 \\
$^{29}$Si & 5.84E-03 & 6.64E-03 & 8.05E-03 & 9.79E-03 & &
$^{56}$Fe & 1.35E-01 & 1.03E-01 & 1.48E-01 & 7.22E-02 \\
$^{30}$Si & 1.06E-02 & 9.05E-03 & 8.76E-03 & 7.72E-03 & &
$^{57}$Fe & 5.45E-03 & 4.19E-03 & 7.80E-03 & 3.31E-03 \\
$^{31}$P  & 1.22E-03 & 1.24E-03 & 1.25E-03 & 1.15E-03 & &
$^{58}$Fe & 2.72E-08 & 1.62E-08 & 1.08E-08 & 3.88E-09 \\
$^{32}$S  & 9.60E-02 & 7.65E-02 & 5.73E-02 & 2.80E-02 & &
$^{59}$Co & 6.57E-04 & 4.62E-04 & 4.14E-04 & 1.28E-04 \\
$^{33}$S  & 4.30E-04 & 3.55E-04 & 2.87E-04 & 1.57E-04 & &
$^{58}$Ni & 8.37E-03 & 1.03E-02 & 9.84E-03 & 8.98E-03 \\
$^{34}$S  & 4.56E-03 & 3.84E-03 & 2.85E-03 & 1.44E-03 & &
$^{60}$Ni & 9.67E-03 & 6.37E-03 & 7.85E-03 & 2.34E-03 \\
$^{36}$S  & 7.28E-06 & 1.30E-06 & 9.88E-07 & 5.56E-07 & &
$^{61}$Ni & 5.32E-05 & 1.00E-04 & 2.73E-04 & 1.74E-04 \\
$^{35}$Cl & 1.65E-04 & 1.20E-04 & 1.09E-04 & 5.99E-05 & &
$^{62}$Ni & 4.27E-04 & 1.31E-03 & 2.44E-03 & 2.27E-03 \\
$^{37}$Cl & 3.39E-05 & 2.43E-05 & 1.87E-05 & 8.30E-06 & &
$^{64}$Ni & 1.75E-09 & 3.93E-11 & 2.45E-12 & 2.43E-13 \\
$^{36}$Ar & 2.21E-02 & 1.54E-02 & 1.06E-02 & 4.70E-03 & &
$^{63}$Cu & 6.61E-06 & 1.10E-05 & 1.22E-05 & 4.92E-06 \\
$^{38}$Ar & 1.60E-03 & 1.34E-03 & 9.07E-04 & 3.67E-04 & &
$^{65}$Cu & 1.78E-06 & 8.56E-06 & 5.36E-06 & 1.72E-06 \\
$^{40}$Ar & 1.15E-07 & 1.57E-08 & 1.23E-08 & 6.28E-09 & &
$^{64}$Zn & 8.62E-05 & 3.23E-05 & 1.60E-04 & 1.51E-05 \\
$^{39}$K  & 9.83E-05 & 7.63E-05 & 7.10E-05 & 2.49E-05 & &
$^{66}$Zn & 1.78E-06 & 1.78E-05 & 4.35E-05 & 5.85E-05 \\
$^{41}$K  & 8.63E-06 & 5.76E-06 & 4.68E-06 & 1.80E-06 & &
$^{67}$Zn & 1.06E-07 & 9.58E-07 & 3.10E-07 & 2.40E-07 \\
$^{40}$Ca & 2.48E-02 & 1.63E-02 & 9.70E-03 & 4.15E-03 & &
$^{68}$Zn & 1.82E-06 & 1.27E-06 & 9.38E-08 & 1.33E-07 \\
$^{42}$Ca & 5.68E-05 & 3.94E-05 & 2.58E-05 & 9.59E-06 & &
$^{70}$Zn & 2.20E-13 & 1.56E-12 & 2.19E-12 & 1.71E-13 \\
\enddata
\end{deluxetable}

\clearpage

TABLE 6:
Nucleosynthesis products ($M_\odot$) of Hypernovae (and Normal supernovae)
before the onset of Radioactive Decays for various explosion energies.
The progenitor model is the 16$M_{\odot}$ He core model.

\vspace{.5cm}

TABLE 7:
Nucleosynthesis products ($M_\odot$) of Hypernovae (and Normal supernovae)
before the onset of Radioactive Decays for various explosion energies.
The progenitor model is the 10$M_{\odot}$ He core model.

\vspace{.5cm}

TABLE 8:
Nucleosynthesis products ($M_\odot$) of Hypernovae (and Normal supernovae)
before the onset of Radioactive Decays for various explosion energies.
The progenitor model is the 8$M_{\odot}$ He core model.

\vspace{.5cm}

TABLE 9:
Nucleosynthesis products ($M_\odot$) of Hypernovae (and Normal supernovae)
before the onset of Radioactive Decays for various explosion energies.
The progenitor model is the 6$M_{\odot}$ He core model.

\vspace{.5cm}

Tables 6 - 9 are available at

\vspace{.2cm}
http://www.astron.s.u-tokyo.ac.jp/~nakamura/research/papers/nakamuratab.ps.gz

\tablenum{10}
\begin{deluxetable}{ccrrrr}
\footnotesize
\tablecaption{Major radioactive elements ($M_\odot$)
of hypernovae (and normal supernovae)
for various explosion energies.
\label{doneisohn}}
\tablewidth{0pt}
\tablehead{\colhead{Progenitor}&
\colhead{{$E$ ($\times$ $ 10^{51}$ ergs)}}
&\colhead{100}&\colhead{30}&\colhead{10}&\colhead{1} \\
\colhead{mass($M_{\odot}$)}&\colhead{species}
}
\startdata
16 &$^{26}$Al & 3.57E-05 & 4.50E-05 & 4.39E-05 & 5.27E-05 \\\
   &$^{41}$Ca & 4.25E-05 & 2.70E-05 & 2.13E-05 & 2.42E-05 \\\
   &$^{44}$Ti & 3.24E-03 & 1.85E-03 & 1.89E-03 & 1.53E-04 \\\
   &$^{56}$Ni & 8.45E-01 & 6.92E-01 & 6.78E-01 & 2.35E-01 \\\
   &$^{57}$Ni & 3.65E-02 & 2.86E-02 & 2.87E-02 & 7.06E-03 \\\hline
10 &$^{26}$Al & 6.81E-06 & 2.17E-05 & 1.89E-05 & 1.66E-05 \\\
   &$^{41}$Ca & 2.75E-05 & 2.04E-05 & 1.84E-05 & 3.07E-05 \\\
   &$^{44}$Ti & 3.16E-03 & 6.45E-04 & 1.22E-03 & 1.81E-04 \\\
   &$^{56}$Ni & 7.14E-01 & 4.65E-01 & 5.63E-01 & 1.97E-01 \\\
   &$^{57}$Ni & 3.55E-02 & 1.70E-02 & 2.93E-02 & 8.84E-03 \\\hline
 8 &$^{26}$Al & 2.08E-03 & 5.56E-03 & 8.34E-03 & 1.23E-02 \\\
   &$^{41}$Ca & 1.45E-05 & 1.03E-05 & 6.19E-06 & 2.02E-06 \\\
   &$^{44}$Ti & 9.82E-04 & 7.35E-04 & 6.34E-04 & 6.69E-05 \\\
   &$^{56}$Ni & 2.37E-01 & 1.68E-01 & 1.98E-01 & 6.09E-02 \\\
   &$^{57}$Ni & 9.02E-03 & 6.52E-03 & 8.98E-03 & 1.70E-03 \\\hline
 6 &$^{26}$Al & 5.33E-04 & 2.42E-03 & 4.08E-03 & 6.73E-03 \\\
   &$^{41}$Ca & 8.60E-06 & 5.75E-06 & 4.67E-06 & 1.80E-06 \\\
   &$^{44}$Ti & 4.18E-05 & 3.27E-04 & 7.01E-04 & 8.81E-05 \\\
   &$^{56}$Ni & 1.35E-01 & 1.03E-01 & 1.48E-01 & 7.22E-02 \\\
   &$^{57}$Ni & 5.44E-03 & 4.18E-03 & 7.79E-03 & 3.30E-03 \\\
\enddata
\end{deluxetable}

\tablenum{11}
\begin{deluxetable}{ccrrrr}
\footnotesize
\tablecaption{Abundance ratios of iron-peak elements
for various explosion energies.
\label{tab:arip}}
\tablewidth{0pt}
\tablehead{\colhead{Progenitor}&
\colhead{$E$ ($\times$ $10^{51}$ ergs)}
&\colhead{100}&\colhead{30}&\colhead{10}&\colhead{1} \\
\colhead{mass($M_{\odot}$)}&\colhead{ratios}
}
\startdata
16 &[Cr/Fe] &   0.191  &   0.054  &  -0.096  &   0.116  \\\
   &[Mn/Fe] &  -0.273  &  -0.366  &  -0.492  &  -0.159  \\\
   &[Co/Fe] &  -0.066  &  -0.347  &  -0.333  &  -0.598  \\\
   &[Ni/Fe] &   0.313  &   0.206  &   0.253  &   0.073  \\\hline
10 &[Cr/Fe] &   0.114  &   0.130  &   0.160  &   0.072  \\\
   &[Mn/Fe] &  -0.302  &  -0.219  &  -0.353  &  -0.066  \\\
   &[Co/Fe] &  -0.022  &  -0.495  &  -0.130  &  -0.301  \\\
   &[Ni/Fe] &   0.374  &   0.064  &   0.573  &   0.450  \\\hline
8  &[Cr/Fe] &   0.183  &   0.286  &   0.090  &   0.179  \\\
   &[Mn/Fe] &  -0.148  &  -0.114  &  -0.364  &  -0.209  \\\
   &[Co/Fe] &   0.138  &   0.007  &  -0.273  &  -0.575  \\\
   &[Ni/Fe] &   0.357  &   0.348  &   0.173  &   0.025  \\\hline
6  &[Cr/Fe] &   0.221  &   0.228  &   0.546  &  -0.014  \\\
   &[Mn/Fe] &  -0.123  &  -0.102  &  -0.459  &  -0.427  \\\
   &[Co/Fe] &   0.221  &   0.184  &  -0.013  &  -0.210  \\\
   &[Ni/Fe] &   0.332  &   0.437  &   0.340  &   0.484  \\\
\enddata
\end{deluxetable}

\end{document}